\documentclass{aa}
\usepackage{graphicx}

\newcommand{\lesssim}{\mathrel{\hbox{\rlap{\hbox{\lower4pt\hbox{$\sim$}}}\hbox{$
<$}}}}
\newcommand{\gtrsim}{\mathrel{\hbox{\rlap{\hbox{\lower4pt\hbox{$\sim$}}}\hbox{$>
$}}}}

\newcommand{\Lx}{L_\mathrm{x}}
\newcommand{\nel}{n_\mathrm{e}}
\newcommand{\me}{m_\mathrm{e}}
\newcommand{\mpr}{m_\mathrm{p}}
\newcommand{\mx}{m_\mathrm{x}}
\newcommand{\mv}{m_\mathrm{v}}

\begin{document}
\title{INTEGRAL observations of 
SS433, a supercritically accreting microquasar with hard spectrum
\thanks
{Based on observations with INTEGRAL, an ESA project with instruments
and science data centre funded by ESA member states (especially the
PI countries: Denmark, France, Germany, Italy, Switzerland, Spain),
Czech Republic and Poland, and with the participation of Russia and the USA.}
}

\author{A.M.Cherepashchuk\inst{1}, R.A.Sunyaev\inst{2,3}, E.V. Seifina\inst{1},
I.E.Panchenko\inst{1}, S.V. Molkov\inst{2}, K.A.Postnov\inst{1} 
}

\offprints{cher@sai.msu.ru}

\institute{Sternberg Astronomical Institute, Universitetskij pr, 13, 119992 Moscow, Russia
          \and 
          Space Research Institute, Russian Academy of Sciences,
              Profsoyuznaya 84/32, 117997 Moscow, Russia
          \and Max-Planck-Institut f\"ur Astrophysik,  Karl-Schwarzschild-Str. 1
              Postfach 1317 D-85741 Garching, Germany 
}

\date{Received  2003; accepted 2003}

\abstract{
Observations of SS433 by INTEGRAL carried out in March -- May 2003 
are presented. SS433 is evidently detected on the INTEGRAL images
of the corresponding sky region in the energy bands 25-50 and 50-100 
keV. The precessional variability of the hard X-ray flux is clearly seen.
The X-ray eclipse caused by the binary orbital motion 
is also detected. A possible origin of
the hard continuum is briefly discussed.  
\keywords{X-ray: binaries -- observations}
}
\titlerunning{INTEGRAL observations of SS433}

\authorrunning{Cherepashchuk et al.}

\maketitle

\section{Introduction}
\label{intro}

SS433 is a unique galactic X-ray binary system --  a microquasar
with mildly relativistic ($v=0.26c$), narrow-collimated (opening angle
$\theta \sim 1^o$, Namiki et al 2003), precessing jets (Margon 1984; Cherepashchuk 1981, 1988;
Crampton \& Hutchings 1981). The system exhibits three photometric
and spectral periodicities related to precession ($P_\mathrm{prec}=162^d.5$),
orbital ($P_\mathrm{orb}=13^d.082$) and nutation ($P_\mathrm{nut}=6^d.28$)
periods (Goranskii
et al. 1998). Recent spectral observations (Gies et al. 2002a)
revealed the presence of absorption lines in the spectrum of the
optical A ($\sim$ A7Ib) supergiant companion. The observed orbital Doppler
shifts of the optical star absorption lines correspond to the mass
ratio of relativistic ($\mx$) and optical ($\mv$) component 
$q=\mx/\mv=0.57\pm 0.11$
and masses $\mv=(19\pm7) M_\odot$, $\mx=(11\pm 5) M_\odot$. 
Similar characteristics were
derived from optical light curves of SS433 (Antokhina \& Cherepashchuk
1987).

Recent infrared spectroscopy (Fuchs et al. 2002) indicated an
enhanced helium abundance in the matter outflowing from the optical
star into the superaccreting precessing accretion disc. This confirmed
the evolutionary status of SS433 as a massive X-ray binary at an advanced
evolutionary stage (Cherepashchuk 1981, 1988) with supercritical
accretion (Shakura \& Sunyaev 1973). The new spectroscopic data (Gies
et al. 2002a) may indicate that the observed large width of X-ray
eclipse in SS433 can be due to collision stellar winds of the binary
components (Cherepashchuk et al. 1995). So it is
established now that SS433 is a microquasar with a black hole (BH),
supercritical precessing accretion disc and precessing quasi-stationary
relativistic jets.

It is well known that X-ray spectra of BH 
X-ray binaries, in contrast to accreting neutron stars, show power-law
tails extending up to many hundreds keV (Sunyaev et al. 1991a,b). In
supercritical accretion regime the thermalisation of
X-ray emission within the  optically thick wind outflowing from the
supercritical accretion disk is expected to cut off the spectrum of an accreting
BH at comparatively low energies, of order of tens keV. At the
same time, geometrical factors, such as the tunnel swept by the
relativistic jets in the disc wind, leave the hope to observe the
possible hard X-ray component from SS433 at least in the precession
phases when the disk is observed mostly face-on (under 
the inclination angle $\sim 30^o$). 

In this paper we report on the searches for hard X-ray component
from the supercritically accreting BH in SS433 using the 
dedicated INTEGRAL observations carried out in May 2003.
We also use the INTEGRAL data on SS433 obtained during 
observations of Aql X-1 (Molkov et al. 2003) 
in March 2003.  

Hard  X-ray emission of SS433 had been observed before by
RXTE (Kotani et al. 2002).
The IBIS detector onboard the INTEGRAL
satellite offers a unique possibility of studying SS433 in
hard X-rays up to hundreds keV (for more detail on INTEGRAL scientific 
payload and mission description see Winkler 1996, 1999).

\section{Observations and data reduction}
\label{observ}

The international gamma-ray observatory INTEGRAL was successively launched 
to its orbit with a Russian rocket PROTON from the Baikonur cosmodrome on
Oct 17, 2002. The scientific payload includes 
instruments designed for the investigation of cosmic
sources in a wide energy band 3-10000 keV: 1) the IBIS telescope, consisting
of the two detectors ISGRI and PICsIT working in the energy band 15-10000 keV,
and allowing sources localization with an  accuracy of about 30\arcsec;
2) the spectrometer SPI, which works in an energy band 15-8000 keV and is designed
for an accurate spectroscopy with the energy resolution
$E/\delta E\simeq 500$ (at 1 MeV); 3) the X-ray monitor JEM-X with an
effective energy range 3-35 keV. All telescopes are using the principle of
coding aperture. The full field of view (FOV) of IBIS telescope, data from which is 
used in this work is $25^o\times 25^o$ with fully coded FOV 
$9^o\times 9$ (FCFOV).

SS433 has been observed by  INTEGRAL in two sets on 11,23,30 March, 6,13 April and 
May 3-9 and 11,
2003, with a total duration of about 500 ks. At this time, SS433 was
seen maximally face-on which favored observations of the deepest parts
of the ``tunnel'' worked out by the jets in the disk wind. On May 3, the
accretion disk was nearly in  front of the optical star and on May
11 it was partially shielded by the optical companion. So these observations
allowed us to detect both the total X-ray emission from the jets and
the X-ray emission from jets with subtracted central parts shielded by the
optical star. Consequently, we can study the ``pure'' spectrum of the
very central jet parts close to the BH. Here 
we present a preliminary quick-look analysis of May 3-8 observations when
the optical star was located behind the accretion disc. These observations
relate to the total emission of the X-ray jet, including its central
part. 

In addition, we have used the INTEGRAL data on SS433 obtained during
observations of Aql X-1 (Molkov et al. 2003) in March-April 2003
to study its hard X-ray 
long-term variability, motivated by the precessional motion observed 
in SS433.  

Calibrations of instruments of INTEGRAL observatory is still not
finished, and right now it is not possible for us to make full 
analysis of data of all 
telescopes. Therefore, our paper focuses on  
the data from the IBIS telescope which has the 
largest effective area. 

For data reduction, we used the standard software package OSA 1.0,
 distributed by INTEGRAL Science Data Center
(ISDC, http://isdc.unige.ch). This version 
does not allow us 
to make a full spectral and timing analysis of data, therefore 
for the preliminary spectral analysis of SS433 we used 
a big set of public calibration observations of the Crab nebula obtained 
in February 2003 (revolutions 39--45). Taking into account the present
uncertainties of the calibrations leading to
strong dependence of the reconstructed source intensity on its distance 
from the center 
of FOV (obtained from the Crab nebula analysis), especially 
in low channels of the ISGRI detector, in our subsequent analysis
we used only the data of observations
when SS433 was inside FCFOV, where the intensity of the source should be
relatively stable. The count rates for the the Crab nebula which were used for the calibration are 
131 counts/s and 89 counts/s in 25-50 and 50-100 keV energy bands, 
respectively.

A reconstruction of the sky regions was made using May 2003 data 
with the total exposures of ISGRI detector $\sim 450$ ks (orbits 67, 68, 69,
70)
for the image reconstruction and $\sim 200$ ks (orbits 67, 68) 
for spectral analysis.

\section{Results}

\subsection{Position accuracy of IBIS and the SS433 image shape}
\label{IBIS_am}


The accuracy of object positions obtained by IBIS has been tested using
the four identified sources -- SS433, Aql X-1, Ser X-1 and GRS 1915+105.
The result 
is that there is a 
systematic shift in the sources' position determination of $\sim 3.9'$
with dispersion $\sim 2'$ 
The systematic shift is mostly along the RA coordinate.
So the X-ray source coincides with the optical position 
of SS433  within the $2'$ error. 
The position accuracy can be ameliorated in future calibrations. 

\begin{figure}
\centering    
\includegraphics[width=\columnwidth]{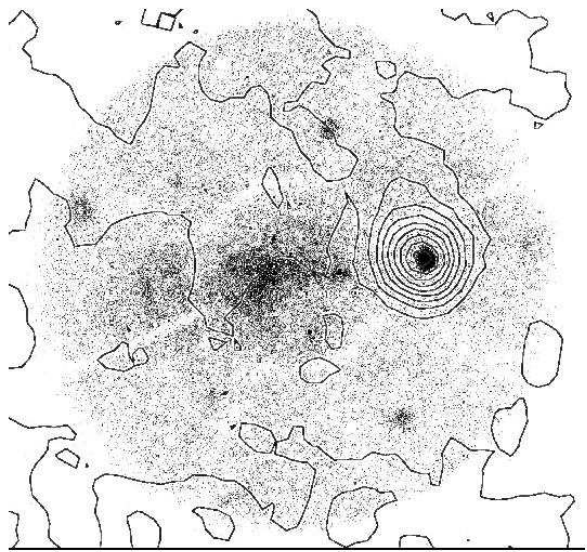}
\caption{Comparison of the ROSAT and IBIS images of SS433. Nothing except the 
central source is detected by IBIS (isolines are IBIS counts per pixel.)}
\label{rosat}
\end{figure}

On the IBIS images, SS433 appears point-like   
with a radius of $\sim 4'$, 
without any significant deviation of isophots from the circular form.  
Any extended details such as 
jets or lobes are not detected. For comparison, in  
Fig.~\ref{rosat} we plot the ROSAT image of SS433 superimposed on
the IBIS 25-50 contours (in counts per pixel). The extended X-ray lobes seen 
on the ROSAT image are not detected 
on the IBIS/ISGRI images by the existing software possibly 
due to insufficient sensitivity and uncertainties
in the background subtraction, ``hot'' pixels tackling procedure, etc.  

\begin{figure}
\centering 
\includegraphics[width=\columnwidth]{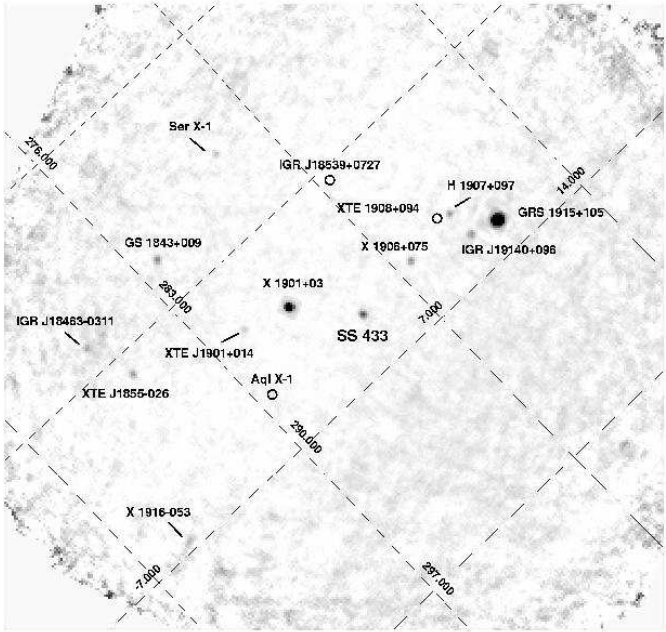}
\caption{Map of the sky around SS433 (in counts per second per pixel) obtained by IBIS/ISGRI telescopes on 
May 2003 in 25-50 keV energy band. The coordinate grid is plotted in the J2000 equatiorial system.
}
\label{ibis25}
\end{figure}

\begin{figure} 
\centering 
\includegraphics[width=\columnwidth]{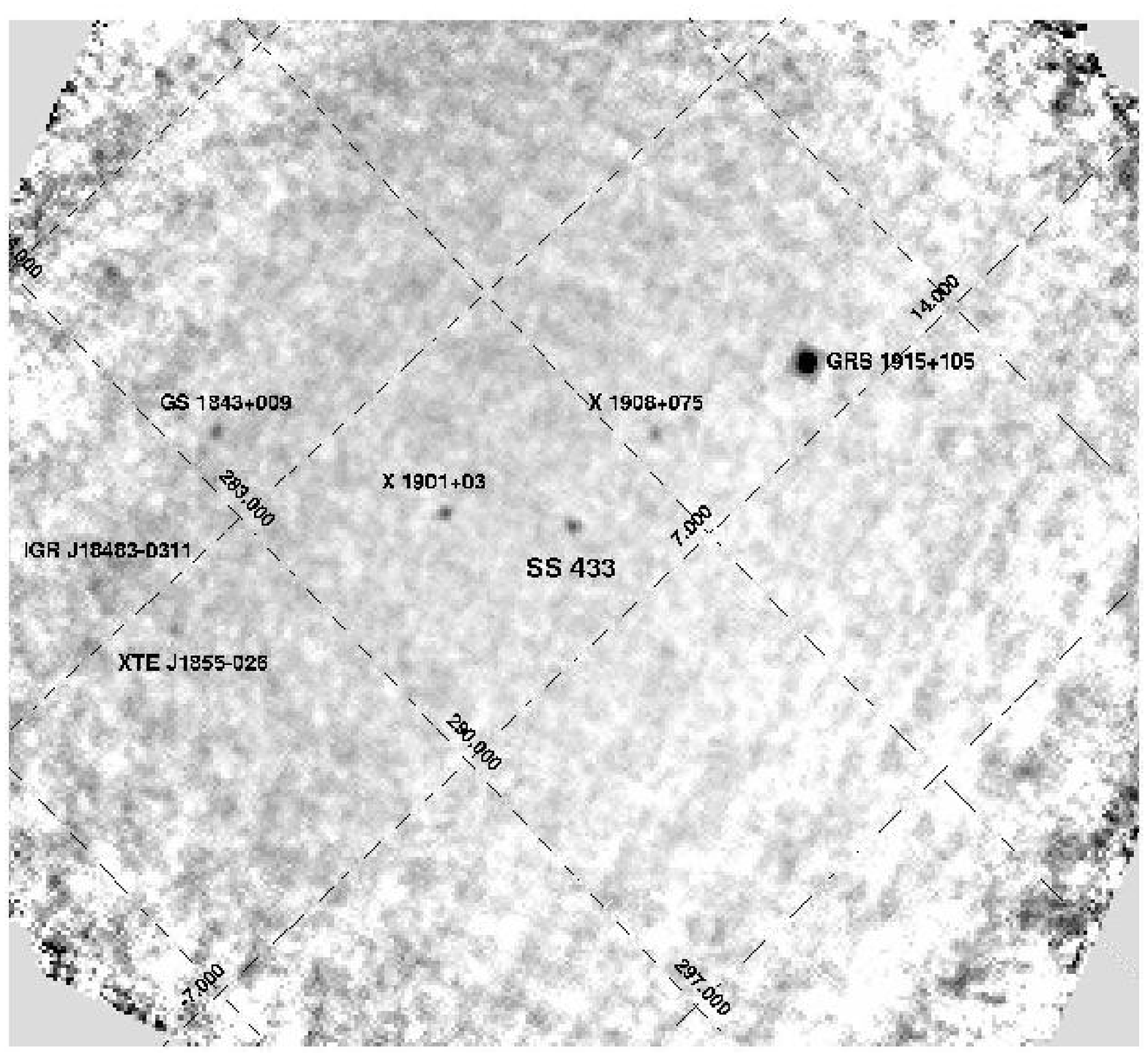}
\caption{Map of the sky around SS433 (in counts per second per pixel) obtained by IBIS/ISGRI telescopes on 
May 2003 in 50-100 keV energy band. The coordinate grid is plotted in the J2000 equatiorial system.}
\label{ibis50}
\end{figure}

SS433 is distinctly  seen in the IBIS FOV at 25-50 keV (Fig. \ref{ibis25})
and  50-100 kev (\ref{ibis50}) energy band,
along with the famous microquasar GRS1915+105 hosting a sub-critically
accreting BH. The very fact of the good seeing of SS433 up
to 100 keV evidences for important role of the geometrical factor
(precessing accretion disk) and may suggest the presence of
relativistic particles in the vicinity of the supercritically accreting
BH (see below).

\subsection {The hard X-ray luminosity of SS433}

SS433 is known to be a moderately bright X-ray source in the  
2-10 keV energy band with $\Lx\sim 10^{35} - 10^{36}$ erg/s 
(Marshall et al. 2002, Namiki et al. 2003) which
is mainly due to jet emission. The luminosity of SS433
in the hard X-rays as inferred from the May 3 INTEGRAL 
observations is $\Lx\sim 3\times 10^{35}$
erg/s (25-50 keV) and $\Lx\sim 1.2\times 10^{35}$ 
erg/s (50-100 keV), assuming the distance of $4.85$~kpc 
(Vermeulen et al 1993)
Thus, the total X-ray luminosity of SS433 is less than that of
classical high-mass X-ray binaries with OB-supergiants (e.g.,
Cherepashchuk et al. 1996).

\subsection{Hard X-ray variability of SS433}

The X-ray flux (2-10 keV) from SS433 is observed by RXTE to be variable
on timescales 100-1000 s (Kotani et al. 2002). This timescale
corresponds to Keplerian frequencies at distances from 10 $M_\odot$ BH
where the 1000 km/s disk wind is formed. Hard X-rays from the jet could
also have such a variability, so we tried to analyze our data to find
it. The hard X-ray variability formed at the very basement of the jets
provides information on the jet production and collimation.  We 
found evidence neither for periodic pulsations on timescales 400-4000 s, nor
quasiperiodic oscillations (typical for subcritically accreting BH). 

\begin{figure} 
\centering 
\includegraphics[width=\columnwidth]{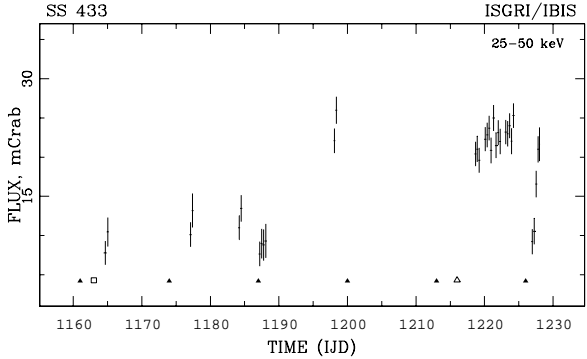}
\caption{Light curve of SS433 (IBIS/ISGRI telescope, 25-50 keV)
on March - May 2003. IJD (INTEGRAL Julian Date) = MJD - 51544.
The filled triangles indicate optical eclipse minima according to
Goranskii et al 1998. The white triangle and square indicate the 
precessional phases of face-on and cross-over respectively.
}\label{lcurve}
\end{figure}

The light curves of SS433 from the IBIS/ISGRI detector exhibit precessional
and orbital variability in the 25-50 keV band. The variations of the flux from 7
to 25 mCrab are clearly detected in the pointings of March 11, 23, 30 and
April 6, 13 as well as on May 3-9, 11. The precessional variability is traced
from the cross-over phase (on March 8), when the disk is seen edge-on and
the X-ray flux is minimal, up to the maximum phase (on May 1), when the disk is
maximally open to the observer. The flux changes more than two times in magnitude with the
precession phase in the 25-50~keV band (see Fig. \ref{lcurve}). The
amplitude of the precessional variability in the 2-12~keV band 
as inferred from RXTE ASM data (Gies et al., 2002b)  
is smaller, $\sim 40\%$.

The pointings of May 9 and 13, 2003 show the decrease of the flux  from
$25$ down to $7$ mCrab, which is apparently due to an X-ray
eclipse in the primary orbital minimum when the ``normal'' star is in front
of the compact source (on May 11). The eclipse phase agrees with 
the ephemeris given by Goranskii et al (1998) 
calculated using optical data only. 
The X-ray eclipse egress is clearly seen as a sharp flux
increase. Though the observation sequence is not continuous,
the X-ray eclipse duration is estimated to be $\approx 2$ days. 
Note that the X-ray flux exhibits essential decrease in all the observed
moments of the expected optical eclipses (shown as black triangles in Fig.
\ref{lcurve}), most clearly on April 2 and May 11.
The eclipses in the 25-50~keV band  are deep and sharp, which possibly 
indicates that the eclipsed emission originates from the compact source itself 
or from the inner parts of the jets.

\begin{figure}
\centering    
\includegraphics[width=0.8\columnwidth,angle=270]{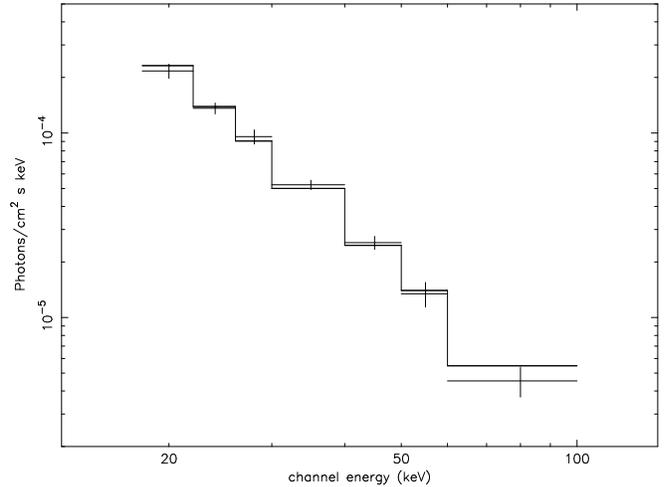}
\caption{The spectrum of SS433 (IBIS telescope, ISGRI mode)
on May 3, 2003. The calibration is made with the use of the Crab
nebula observations, as the standard data processing package does not provide
adequate calibration tools.  
}\label{spectrum}
\end{figure}

\subsection{Hard X-ray spectrum of SS433}

\begin{figure}
\centering    
\includegraphics[width=0.7\columnwidth,angle=270]{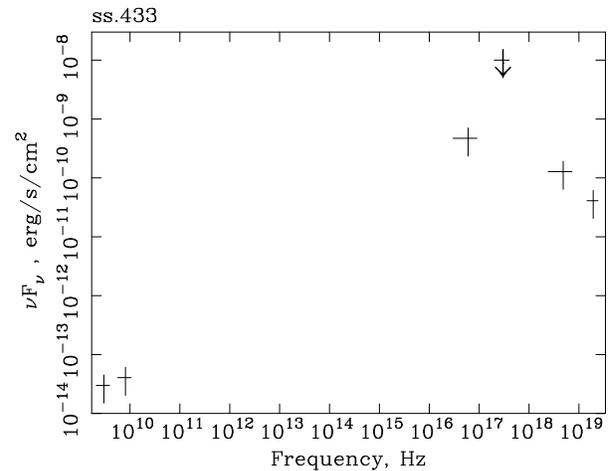}
\caption{The total energy spectrum of SS433 in units $\nu F_\nu$
from radio to hard X-rays. The radio, UV and optical data is taken from
Gies et al, 2002 and Fender et al, 2000. The interstellar extinction
is taken into account ($A_\mathrm{V}=7.5^m$).
}\label{all_sp}
\end{figure}

Despite the preliminary character of our analysis, the hard X-ray 
spectrum of SS433 appears relatively flat from 20 keV to 100 keV
(Fig. \ref{spectrum})  with the power-law photon index $\alpha\sim -2$. Such spectra
are typical for accreting BHs (Sunyaev et al. 1991a,b).

The most straightforward explanation of this spectrum is thermal
Comptonisation of soft photons emitted by the inner parts of the
disc on relativistic electrons in a hot non-thermal corona above
accretion disc. Such coronas are thought to exist in BH
X-ray binaries in the low/hard states and disappear (due to effective
Compton cooling) when accretion luminosity increases (at the high/soft
state). So it is hard to imagine how such a corona could survive
a supercritical accretion transition in the case of SS433. 
It is also problematic to observe directly the hard photons generated in 
a supercritical accretion disk due to photon trapping effects 
(e.g. Ohsuga et al. 2003).

Another explanation could be particle acceleration in Poynting-dominated jets
in the vicinity of the BH. The maximum Lorentz-factor of particles in
a Poynting dominated outflow is (Michel 1969)  
\begin{equation}
\gamma_\mathrm{max}\sim
\sigma^{1/3} 
\end{equation}
where $\sigma=B^2/(4\pi \nel\me c^2)$  is plasma
magnetization parameter ($B$-- magnetic field strength at the jet base,
$\nel$ is the electron density). For electron-proton jet launched by a
relativistic rotator with circular frequency $\Omega$ the magnetization
parameter can be rewritten as (see also Camenzind 1990)
\begin{equation}
\sigma=\frac{\mpr}{\me}\frac{B^2 R^2}{\dot M_\mathrm{j} c}\left(\frac{\Omega
R}{c}\right)^2 
\end{equation} 
where $\dot M_\mathrm{j}=4\pi  \nel \mpr R^2 c$ is the mass
ejection rate at the jet base. For estimation, near the BH we can put
$\Omega R/c\sim 1$ and assume a maximum magnetic field from  $B^2/4\pi
\sim \nel \mpr c^2$. Then $\sigma\sim \mpr/\me\sim 2000$ and
$\gamma_\mathrm{max}\sim 12$. Some additional acceleration can occur outside
the fast magnetosonic surface due to a strong photon field ($\Lx\sim
L_\mathrm{edd}$) around the jet (Beskin, Zakamska \& Sol 2003), as the compactness
parameter in this case is $l_\mathrm{a}=(L/R) \sigma_\mathrm{T}/(\me c^3)\sim 10^3-10^4>\sigma$.
This estimate shows that hard (up to 100 keV) X-ray photons can be Compton
upscatterred soft X-ray photons from the inner parts of the disk on mildly
relativistic electrons accelerated in the Poynting dominated jet 
formed close to the BH. We stress here that
actually observed jet outflow $\sim 10^{-6} M_\odot$/yr with $v=0.26 c$ is an
order of magnitude larger than the Eddington-controlled matter flux near BH
and should be formed around and upstream the initial jet involving
non-relativistic disk wind. The observed precession of the jets locked with
disk precession firmly indicates that the most jet outflow in SS433 
is due to the
disk wind. We also note that in the jets, various thermal instabilities can
operate giving rise to their inhomogeneous structure. More detailed analysis
of rapid variability of SS433 in hard X-rays is in progress.

The total spectrum of SS433  from radio to hard X-rays is shown in 
Fig. \ref{all_sp}. Maximum energy release is in the optical 
band, as expected from a supercritically accreting disk in which 
most energy released in the disk is thermalised by the optically thick 
disk wind. We stress again that the presence of hard X-ray emission
from the supercritically accreting disk in SS433 
and its variability with precessional period provides evidence 
for a hollow cone in the polar regions of the disk apparently  
swept-up in the disk wind by precessing jets.

\section{Conclusion}

We  have presented a quick-look analysis of 
SS433 by the IBIS telescope onboard the INTEGRAL gamma-ray observatory. 
The observations revealed for the first time the clear presence of a
variable hard X-ray continuum in the SS433 (up to 100 keV). Hard X-rays
from supercritical accreting disk around a 10 $M_\odot$ BH can
be due to Comptonisation of soft X-ray photons generated in the inner
disk on relativistic electrons accelerated in the central part of the
jet near the BH.

\acknowledgements
Research has made use of data obtained through the INTEGRAL Science Data
Center (ISDC), Versoix, Switzerland, and Russian INTEGRAL Data Center, Space
Research Institute, Moscow, Russia. 
The authors acknowledge N.I.Shakura, A.N.Timokhin, and E.A.Antokhina for 
valuable notes.
The work is partly supported by the Russian Foundation for Basic
Research (projects 00-15-96553 and 00-02-17164) and by the program of
the support of leading scientific schools of Russia.

\end{document}